\documentclass[useAMS,usenatbib,10pt,a4paper,draft=false]{mn2e} % ,draft
\usepackage[utf8]{inputenc}
\usepackage[T1]{fontenc}
\usepackage{aecompl} 
\usepackage{aas_macros}
\usepackage{amssymb,amsmath}
\usepackage{booktabs} % for nice tables
\usepackage{graphicx}
\usepackage{lmodern}
\usepackage{siunitx}
\usepackage{subcaption}
\usepackage{times}
\usepackage{xspace} 
\usepackage[
colorlinks=true, 
linkcolor=blue, 
anchorcolor=blue,
citecolor=blue, 
filecolor=blue, 
menucolor=blue,
urlcolor=blue]{hyperref} 

\newcommand{\Mn}{\ensuremath{{M}_\mathrm{N}}}
\newcommand{\Mt}{\ensuremath{{M}_\mathrm{T}}}
\newcommand{\Mth}{\ensuremath{M_\text{thresh}}}
\newcommand{\Msun}{\ensuremath{\mathrm{M}_\odot}}
\newcommand{\diff}{\ensuremath{\mathrm{d}}}

\newcommand{\vect}[1] {\ensuremath{\mathbf{#1}}}

\newcommand{\hMpc}[1]{\ensuremath{{#1}\,h^{-1} \mathrm{Mpc}}}

\newcommand{\hkpc}[1]{\ensuremath{{#1}\,h^{-1} \mathrm{kpc}}}
\newcommand{\hMsun}[1]{\ensuremath{\num{#1}\,h^{-1} \Msun}}

\newcommand{\go}{\texttt{GOTPM}}
\newcommand{\ie}{\emph{i.e., }}
\newcommand{\Om}{\ensuremath{\Omega_\text{m}}}
\newcommand{\Ol}{\ensuremath{\Omega_\Lambda}}

\begin{document} 

\title[Ecology of haloes I]{% 
  The ecology of dark matter haloes I:\\
  The rates and types of halo interactions
}
\author[B.~L'Huillier, C.~Park and J.~Kim]
{Benjamin~L'Huillier,$^1$ 
  Changbom~Park$^1$
  and 
  Juhan~Kim$^{2}$\thanks{E-mail:lhuillier@kias.re.kr             (BL),
    cbp@kias.re.kr (CP), kjhan@kias.re.kr (JK)}\\ 
  $^1$ School of Physics, Korea Institute for Advanced
  Study, 85 Hoegi-ro, Dongdaemun-gu, Seoul 130-722, Korea\\
  $^2$ Center  for Advanced Computation, Korea  Institute for Advanced
  Study, 85 Hoegi-ro, Dongdaemun-gu, Seoul 130-722, Korea 
} 

\date{Accepted 2015 April 29. Received 2015 April 27; in original form
  2015 March 11 } 

\pagerange{\pageref{firstpage}--\pageref{lastpage}} \pubyear{2015}

\maketitle
 
\label{firstpage}

\begin{abstract} 
Interactions such  as mergers  and flybys play  a fundamental  role in
shaping galaxy morphology.
Using the Horizon  Run 4 cosmological $N$-body  simulation, we studied
the  frequency  and type  of  halo  interactions, and  their  redshift
evolution as a function of  the environment defined by the large-scale
density, pair separation, mass ratio, and target halo mass. 
Most  interactions  happen  at large-scale  density  contrast  $\delta
\simeq 20$,  regardless of  the redshift  corresponding to  groups and
relatively dense part of filaments.
However,  the fraction  of interacting  target is  maximum at  $\delta
\simeq 1000$.
We provide a new empirical fitting  form for the interaction rate as a
function of the halo mass, large-scale density, and redshift.  
We also  report the existence  of two  modes of interactions  from the
distributions  of  mass  ratio  and relative  distance,  implying  two
different physical origins of the interaction.  
Satellite targets lose their mass as they proceed deeper into the host
halo. 
The relative  importance of these  two trends strongly depends  on the
large-scale density, target mass, and redshift.  
\end{abstract}

 
\section{Introduction}

In a  hierarchical $\Lambda$CDM  universe, low-mass haloes  form first
and  merge together  into  a more  massive  one through  gravitational
attraction.  
Interactions might prevail during the halo formation and evolution.
Therefore, the star-formation activity  and the morphology of galaxies
are governed by the halo interactions. 

 \bigskip

The effects of interactions also depend on the environment.
In  dense environments  like clusters,  intracluster interactions  can
suppress star formation by  multiple high-velocity galaxy interactions
within  the cluster  \citep[harassment,][]{1996Natur.379..613M} or  by
tidal stripping \citep{1984ApJ...276...26M, 1990ApJ...350...89B}.  
In addition to gravitational interactions, hydrodynamics also plays an
important       role      with       ram      pressure       stripping
\citep{1972ApJ...176....1G, 2014MNRAS.438..444B, 2014ApJ...781...38C}. 
Starvation  or strangulation  strips  the hot  gas  and quenches  star
formation   on   a   larger   time-scale   \citep{1980ApJ...237..692L,
  2014ApJ...781...38C}. 
  In particular, minor mergers have been  shown to be an efficient way
  to     quench    star     formation,    both     from    theoretical
  \citep[e.g.,     ][]{1994ApJ...425L..13M}      and     observational
  \citep[e.g., ][]{2014MNRAS.440.2944K} work. 
On the  other hand, mergers  have thus been  considered as one  of the
most   efficient   processes   to  shape   galaxies,   turning   blue,
star-forming,  spiral   galaxies  into  red,  dead,   elliptical  ones
\citep{1972ApJ...178..623T, 1978ApJ...219...46L}.  

\bigskip
 
The density--morphology dependence has been studied for decades.  
Rich clusters are dominated by a giant elliptical, and the fraction of
ellipticals    decreases     with    increasing     radial    distance
\citep{1974ApJ...194....1O, 1980ApJ...236..351D, 1984ApJ...281...95P}.  
On the contrary, fields are dominated by blue galaxies \citep 
{2003ApJ...584..210G, 2003MNRAS.346..601G, 2004AJ....128.2677T}.  

\bigskip

\citet{2008ApJ...674..784P} found --- from the SDSS DR4 --- that galaxies
located within  the virial radius  of their nearest neighbour  tend to
have  the  same morphology  as  their  neighbour as  their  separation
decreases,  whereas if  their  separation is  larger  than the  virial
radius,  the   fraction  of   elliptical  decreases   with  increasing
separation. 
Subsequent  studies   at  low   redshifts  \citep{2008ApJ...674..784P,
  2009ApJ...691.1828P},   high   redshift   \citep[$0.4\leq   z   \leq
1.0$][]{2009ApJ...700..791H}      and     low-density      environment
\citep{2014arXiv1406.3868Y} confirmed these results.  
  
\bigskip

The  \citet{1974ApJ...187..425P}  formalism   ---  and  its  extensions
\citep{1991ApJ...379..440B} ---  enabled the  study of the  merger rate
from    dark    matter    only    \citep[e.g.,][]{2006ApJ...652...56B,
  2008MNRAS.386..577F,    2009MNRAS.394.1825F,    2010ApJ...715..342H,
  2010ApJ...719..229G}  or   hydrodynamic  \citep{2002ApJ...571....1M,
  2006ApJ...647..763M,    2012A&A...544A..68L,    2014arXiv1411.2595K,
  2015MNRAS.449...49R, 2015arXiv150205053W} cosmological simulations.  
However, the merger rate still varies by an order of magnitude between
different models \citep{2010ApJ...724..915H}.  
Environmental studies from SAMs \citep[e.g.,][]{2014ApJ...790....7J}
and  $N$-body  simulations \citep[e.g.,][]{2009MNRAS.394.1825F}  have
also  shown  an  enhancement  of  the  merger  rate  in  high-density
regions. 

\bigskip

Observationally,  the merger  rate  is usually  estimated either  from
close-pair  counts   \citep{2002ApJ...565..208P,  2014ApJ...795..157K,
  2014MNRAS.444.3986R, 2015A&A...576A..53L} 
or  by   morphological  signatures   \citep{2003AJ....126.1183C},  and
converted through a merger timescale.  
However, this merger  timescale depends on the mass ratio  and the gas
fractions,  which  may  lead  to  uncertainties  in  the  merger  rate
\citep[e.g.,][]{2010MNRAS.404..590L, 2010MNRAS.404..575L}. 
  Several studies have used visual inspection to study the morphology 
  \citep{2010MNRAS.401.1552D,                     2011MNRAS.410..166L,
    2014MNRAS.437L..41K}. 

 \smallskip
 
$N$-body  simulations of  galaxy  encounters have  shown that  violent
mergers  were  able to  turn  spiral  galaxies into  elliptical  ones,
supported    by     the    random     motions    of     their    stars
\citep[e.g.,][]{1972ApJ...178..623T,       1988ApJ...331..699B,
  1992ApJ...393..484B}. 
The implementation of hydrodynamics  showed that gravitational torques
from  the interacting  pairs  could bring  gas to  the  centre of  the
galaxies \citep{1991ApJ...370L..65B,  1996ApJ...471..115B}, leading to
bursts   of  star   formation  \citep[e.g.,][]{1996ApJ...464..641M,
  2002MNRAS.333..327T, 2007A&A...468...61D, 2014MNRAS.442.1992H}.  

 \smallskip

However, not all interactions end in mergers.  
If the relative  velocity of the interacting galaxies  is high enough,
they will not merge within a dynamical timescales.  
This  class of  interaction,  referred  to as  flyby,  has been  less
studied, although in the recent years its interest has been growing
\citep    [e.g.,]   []    {2012MNRAS.419..411M,   2012ApJ...751...17S,
  2012MNRAS.425.2313T}. 
At low redshifts on high mass scales, flybys becomes as frequent as, or
even more frequent than  mergers \citep{2012ApJ...751...17S}.
\citet{2012MNRAS.419..411M}    used    the    Millennium    simulation
\citep{2005Natur.435..629S}  to  analyse  interacting pairs  that  are
closer than a fixed comoving distance $d_\text{crit}$ at some point of
their existence.  
They  found  that 53\%  of  pairs  never  merge for  $d_\text{crit}  =
\hMpc{1}$, while a lower $d_\text{crit}$ reduces this fraction.  
Therefore,  difficulties  arise in  observational  work,  since a  non
negligible  fraction of  observed pairs  might not  be gravitationally
bound \citep{2012ApJ...751...17S, 2012MNRAS.425.2313T}.  
The  presence of  a hot  gaseous halo  in late-type  galaxies is  also
important  for the  gas transfer  and the  star formation  rate during
distant interactions \citep{hwangpark15}.

 \smallskip

A key factor in understanding galaxy evolution, is to understand which
of these phenomena (close and distant encounters) dominates.  
Distant  interactions  are weaker  than  close  ones,  but occur  more
frequently. 
It is then  important to determine which one is  the most efficient at
shaping galaxy morphology. 
In this study, we used the  large Horizon Run 4 $N$-body simulation to
study the environmental dependency of the interaction rate, regardless
of their final stage (merger or flybys).

In \S~\ref{sec:meth} we describes the  set of simulations used in this
work and introduce our definition of interaction and environment.  
Our   results   about   the   interaction  rate   are   presented   in
\S~\ref{sec:res},  and a  study  of  the distance  and  mass ratio  is
performed in \S~\ref{sec:type}. 
\S~\ref{sec:disc} discusses our results.
Finally, the conclusions are drawn in \S~\ref{sec:ccl}.

\section{Simulation and method}
\label{sec:meth}

We present the Horizon Run 4 simulation used here in \S~\ref{sec:sim},
and detail  the method  we used  to construct  the halo  catalogues in
\S~\ref{sec:catalogue}.  
Interactions are  defined in \S~\ref{sec:inter}  and \S~\ref{sec:dens}
describes the density calculation.

\subsection{Simulations}
\label{sec:sim}
We  used the  Horizon run  4 simulation  \citep{hr4} performed  with a
memory-efficient version of  the \go\ code \citep{2004NewA....9..111D,
  2009ApJ...701.1547K,   2011JKAS...44..217K},   with   a   box   size
$L=\hMpc{3150}$, and $N=6300^3$ particles.  
The simulation used second order Lagrangian perturbation theory (2LPT)
initial  conditions  at  $z_\text{i}   =100$  and  a  WMAP5  cosmology
$(\Omega_\text{b},  \Omega_\text{m},\Omega_\Lambda,h,\sigma_8,n_s)$  =
(0.044, 0.26, 0.74, 0.72, 0.79, 0.96), yielding a particle mass of 
$m_\text{p} \simeq \hMsun{9.02e9}$.  
This starting redshift combined  with 2LPT initial conditions, ensures
an     accurate      mass     function     and      power     spectrum
\citep{2014NewA...30...79L}. 

\subsection{Halo catalogues}

\label{sec:catalogue}
We  detected haloes  from the  simulated particle  distributions using
OPFOF,  a   MPI-parallel  version  of  the   friends-of-friends  (FOF)
algorithm \citep{1985ApJ...292..371D}, with  a standard linking length
of  0.2  mean  particle separations,  and  the  gravitationally-stable
subhaloes  with the  PSB algorithm  \citep{2006ApJ...639..600K}, which
are assumed to host galaxies.  
PSB  structures identified  in an  FoF halo  are considered  satellite
subhaloes, except  the most massive one  which is defined as  the main
halo hosting the main galaxy.
In  the  following,   haloes  will  indistinctly  refer   to  main  or
subhaloes.

The virial mass $M_\text{v}$ of each  subhalo is defined as the sum of
the mass of its member particles. 
The    virial    radius    $R_\text{v}$    is    then    defined    by
\begin{align}
  \label{eq:rvir}
  R_\text{v} &= \left(\frac{3 M_\text{v}}  {4 \pi \Delta_\text {c} (z)
      \rho_\text{c} (z)} \right)^{1/3}, \\
\intertext{where the critical density is defined as} 
\rho_c(z) &= 3H^2(z)/(8\pi G),
\end{align}
and the critical overdensity  at redshift $z$, $\Delta_\text{c}(z)$ is
well  described by  the empirical  \citet{1998ApJ...495...80B} fitting
formula 
\begin{align}
  \label{eq:delta_c}
  \Delta_\text{c}(z) &= 18\pi^2 + 82 x - 39 x^2,\\ 
 \intertext{where} 
  x &= \Omega(z)-1 = -\frac{\Ol}{\Om (1+z)^3 + \Ol}. 
\end{align} 
Note that  this definition of  the virial  radius is in  physical (not
comoving) units.  
However, in the following, we will only use comoving distances, and we
include the Hubble flow $H\vect r$ in the definition of the velocity.

{%
We  defined  target haloes  (T)  as  main  or  subhaloes with  $\Mt  >
\Mth$.
We chose  $\Mth \equiv  \, \hMsun{5e11}$,  corresponding to
the mass of 56 particles. 
This   choice  yields   a   mean  halo   separation  of   \hMpc{5.17},
corresponding   to  SDSS   galaxies  brighter   than  $M_r   =  -19.8$
\citep{2010JKAS...43..191C}.  
}
For  the neighbour  haloes  (N), we  limit our  catalogue  to main  or
subhaloes  with  mass $\Mn  >  q_0  \Mth$, with  $q_0=0.4$,
yielding a minimal number of particles of 23.  
{%
The  choice  of \Mth\  and  $q_0$  are important  for  the
construction of our catalogues.
A lower  \Mth\ will  result into  a larger  target catalogue,  while a
lower $q_0$ enables to probe larger mass ratios.
We  then  chose  $q_0=0.4$  as  a good  trade-off,  yielding  $\Mth  =
\hMsun{5e11}$. 
}
The  mean  separation   of  haloes  in  the   neighbour  catalogue  is
\hMpc{3.97} at $z=0$, corresponding to  that of SDSS galaxies brighter
than $M_r = -18.7$ \citep{2010JKAS...43..191C}.  
Note  that  the target  catalogue  is  a  subsample of  the  neighbour
catalogue. 
The minimal  number of particles  in the PSB  subhaloes is set  to 20,
yielding a  minmal mass  of $\hMsun{1.8e11}$, therefore  our neighbour
catalogue is complete.  
The    target   (neighbour)    catalogue   contains    \num{225406978}
(\num{472635985}) haloes at $z=0$.  
Note  that we  only consider  the  nearest neighbour,  and ignore  the
effects  of  a  second  neighbour,  which  may  become  important  for
satellite pairs in clusters \citep{2013MNRAS.436.1765M}. 
{%
  We discuss this choice in \S~\ref{sec:neighbour}
}

\subsection{Quantification  of  the environment:  large-scale  density
  and distance to the nearest neighbour}
\label{sec:dens}
The environment may be a key factor to define the interaction rate.
Both  large-scale and  local  environment have  been  shown to  affect
galaxy interactions.  
\citet{2012MNRAS.419.2670M}  compared  different  ways to  define  the
environment:  using a  fixed aperture  or  the distance  to the  $n$th
nearest neighbour.  

Another way to  define and quantify the environment is  to compute the
eigenvalues of  the smooth tidal field  \citep{2009MNRAS.396.1815F} or
velocity-shear field \citep{2012MNRAS.425.2049H}.

\begin{figure}%[t!]
  \centering
  \includegraphics[width=\columnwidth]{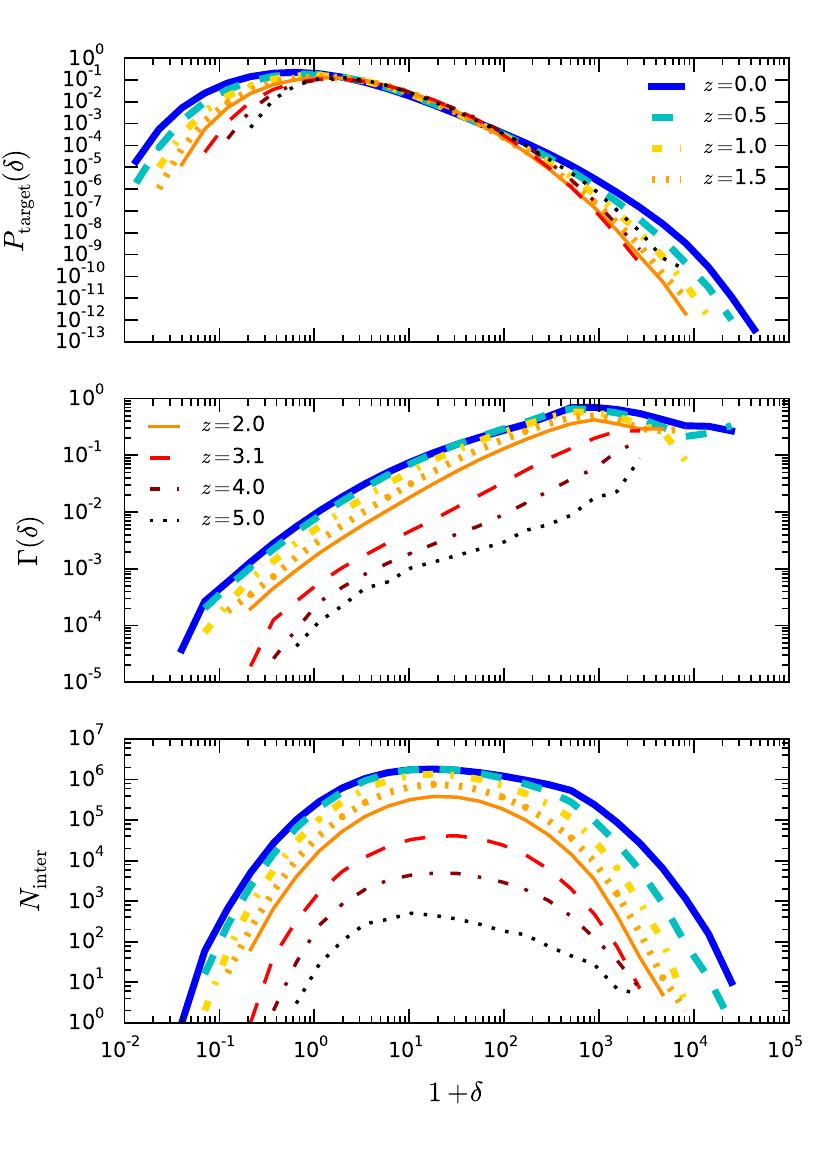} 
  \caption{\label{fig:frho}%
    \emph{Top:} Distribution  of the large-scale  background density
    at the  location of  the target  haloes at  different redshifts.
    \emph{Middle:}  Fraction  of  target haloes  that  encounter  an
    interaction  as  a  function   of  the  large-scale  environment
    $\delta$. 
    \emph{Bottom:} Histogram of interactions per density bin.  
  }  
\end{figure} 

We defined  the background  density as  in \citet{2007ApJ...658..898P}
and \citet{2008ApJ...674..784P}, by 
\begin{equation} 
\rho_{20} \equiv \sum_{i=1}^{20} M_i W(r_i,h),
\end{equation}
where $r_i$ and  $M_i$ are the distance and mass  of the $i$th closest
galaxy  in   the  catalogue,  and   $W$  is  the  SPH   spline  kernel
\citep{1985A&A...149..135M} of smoothing length $h$.  
The latter is searched to enclose  20 neighbours, which has been shown
to be the smallest number yielding an accurate estimate of the density
\citep{2007ApJ...658..898P}.
The adaptive nature of this  technique enables to accurately probe the
density     both     in      high-     and     low-density     regions
\citep{2007ApJ...658..898P}. 
Appendix~\ref{sec:delta} shows the  relation between $\delta$ and the
cosmic web.
 
Given the  large number  of haloes,  we adopted  a set  of oct-sibling
trees build in rectangular grids over the entire volume to efficiently
compute the density, similar to \citet{2004NewA....9..111D}.

The local density is defined by  the mass of the nearest neighbour and
the separation $r$:
\begin{equation}
  \rho_\text{n} \equiv \frac  {3 \Mn} {4\pi r^3 }  = \frac {\Delta_\text{c}
    \rho_\text {c}} {p^3}, 
\end{equation}
and is thus fully described by the ratio of the pair separation to the
virial  radius  of   the  neighbour  at  a   given  redshift  $p\equiv
d/R_\text{vir,n}$.

The mean  density is  the total  mass of the  haloes in  the Neighbour
catalogue
\begin{equation}
  \label{eq:rhobar}
  \bar{ \rho} =\frac 1 V \sum{M_i}.
\end{equation}
This  enables us  to normalise  the density  at different  redshift by
considering 
\begin{equation}
  \delta = \frac {\rho_{20}}{\bar \rho}-1.
\end{equation}
We  find   $\bar{\rho}  (z=0)  =  2.51\times   10^{10}\,h^2\,  \Msun\,
\mathrm{Mpc}^{-3}$.

\begin{figure*}
  \centering
  \begin{subfigure}[t]{\columnwidth}
    \includegraphics[width=\textwidth]{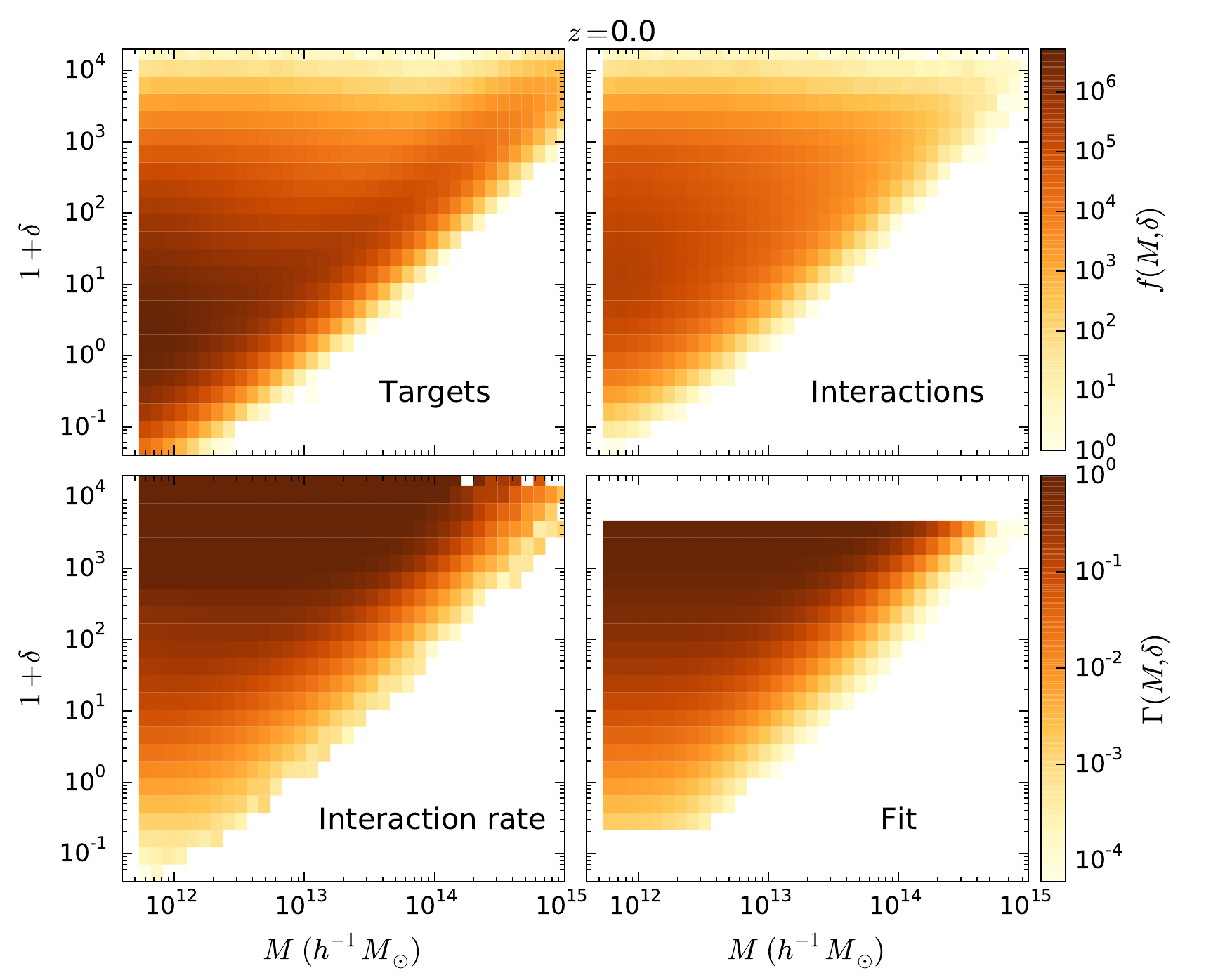}
    \caption{\label{fig:mrhoz0}%
      $z=0$
    }
  \end{subfigure}
  \begin{subfigure}[t]{\columnwidth}
    \includegraphics[width=\textwidth]{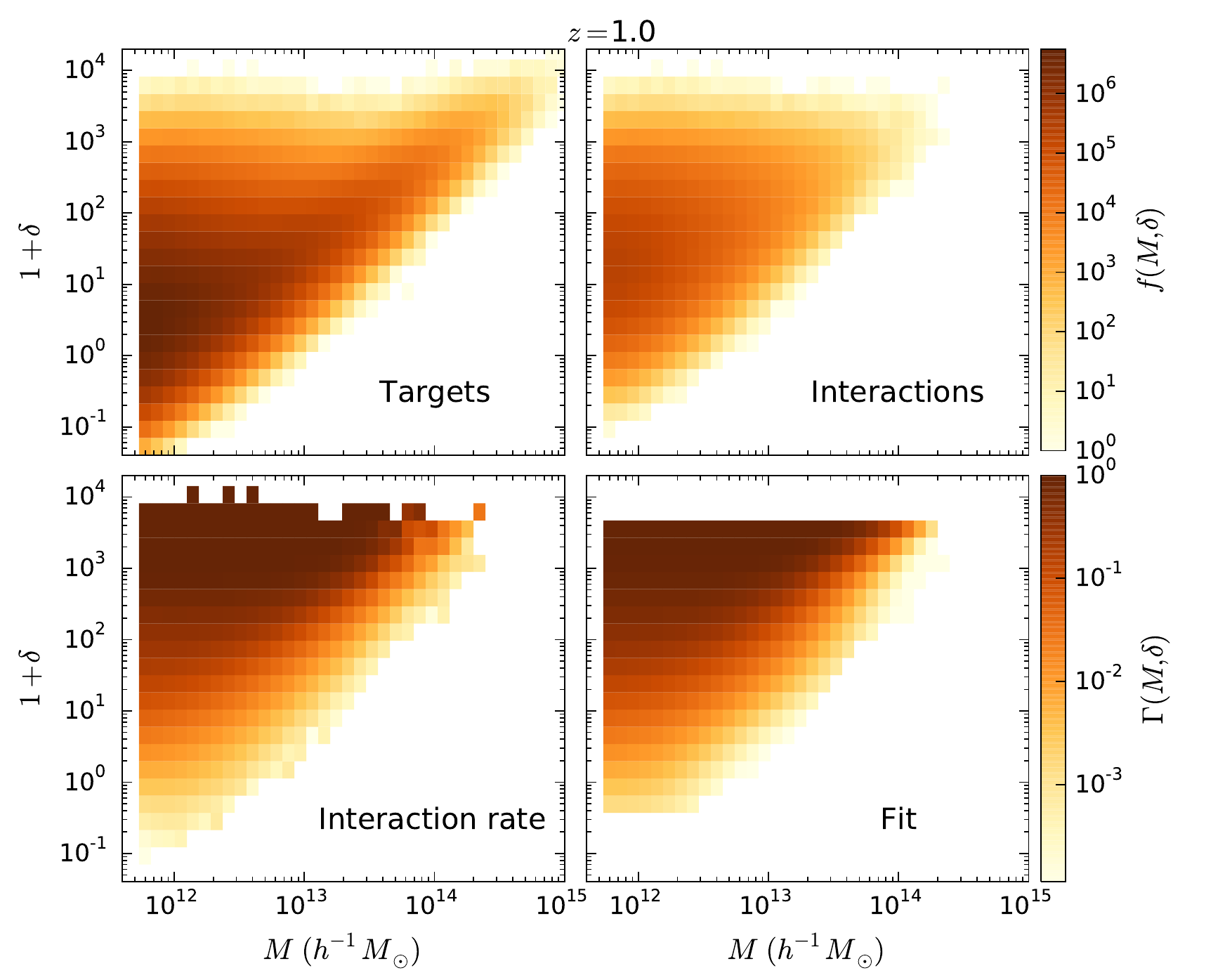}
  \caption{\label{fig:mrhoz1}%
    $z=1$
  }
  \end{subfigure}
  \caption{\label{fig:mrho}%
    Distribution  function  of  the   target  haloes  (top  left)  and
    interacting  target (top  right) as  a  function of  the mass  and
    large-scale   density  at   $z=0$   (\subref{fig:mrhoz0})  and   1
    (\subref{fig:mrhoz1}).  
    Bottom  left: interaction  rate  as  a function  of  the mass  and
    large-scale density. 
    Bottom    right:    fit    to   the    interaction    rate    with
    eq.~\eqref{eq:gamma_m}.  
  }
\end{figure*}

\subsection{Definitions of interactions}
\label{sec:inter}
We define  interacting systems  as target haloes  with mass  $\Mt \geq
\Mth$ located within the virial  radius of a neighbour more
massive than $q_0\Mt$.  
Such target haloes are considered  to be gravitationally influenced by
their neighbour.

\begin{figure}
  \includegraphics[width=\columnwidth]{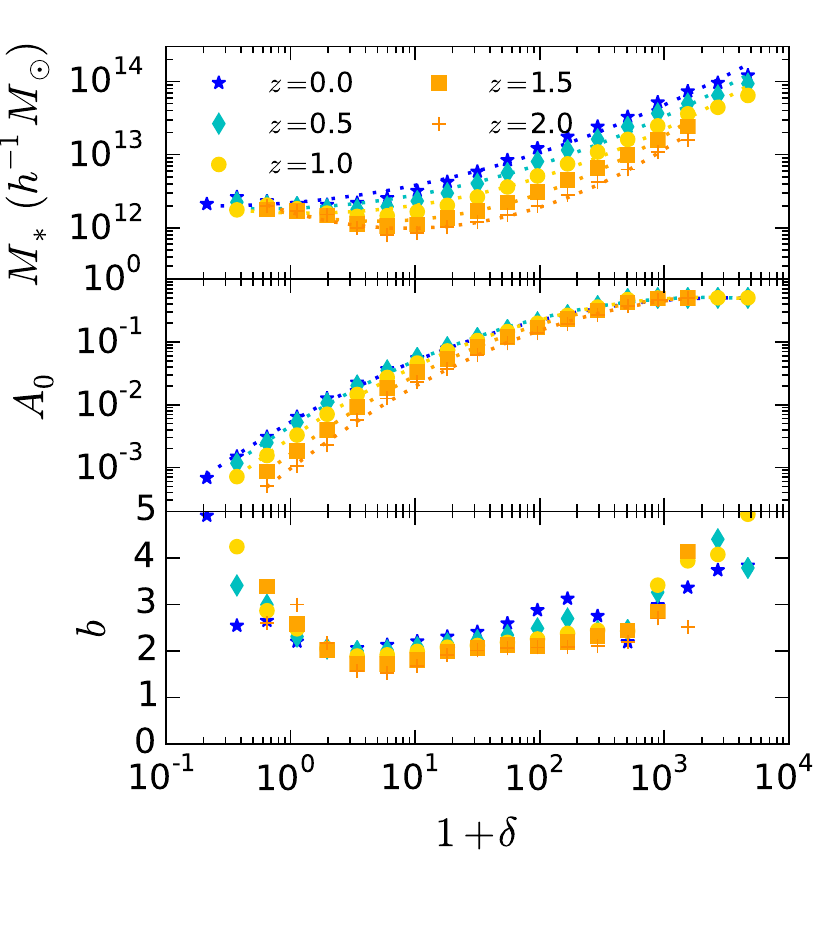}
  \caption{\label{fig:params}%
    Dependence    of   the    parameters   $A_0,    M_*$,   and    $b$
    of equation~\eqref{eq:gamma_m} on $\delta$ and $z$.  
  }
\end{figure} 

We define  the interaction rate  as the  fraction of targets  that are
undergoing an interaction over the total number of target haloes.
This definition  slightly differs  from some  authors, who  define the
merger  rate   as  the   number  of   merger  events   per  descendant
\citep[e.g.,][]       {2008MNRAS.386..577F,       2009MNRAS.394.1825F,
  2010MNRAS.401.2245F}. 
Indeed, one can notice that an interaction may be counted twice if the
mass ratio between two neighbours is between $q_0$ and $1/q_0$, and if
their separation is smaller than both virial radii (which is likely to
happen, since the virial radii of these two haloes are comparable).

\begin{figure}  
  \includegraphics[width=\columnwidth]{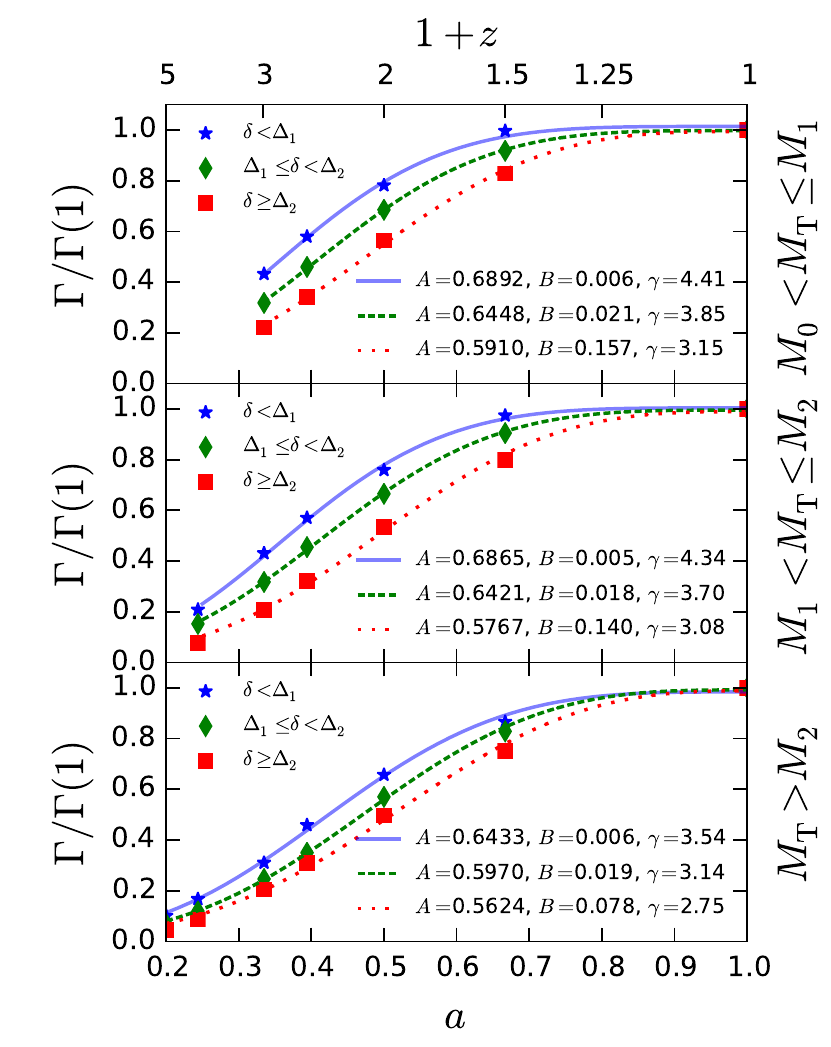}
  \caption{\label{fig:Gamma_z}%
    Interaction rate (normalised to the  final value) as a function of
    the scale factor $a$ for low- (top), intermediate- (middle), and
    high-mass  (top)   bins,  and   for  low-   (stars),  intermediate
    (diamonds), and high-density (squares), and the associated fit.  
  }
\end{figure}

We  will focus  on  several  quantities: the  mass  of  the halo,  its
background density,  the mass  ratio $q  = \Mn/\Mt$  and $p$  the pair
separation divided by the virial radius of the neighbour.  
We  also note  that, in  our  definition, the  mass ratio  $q$ of  the
interaction is  by definition larger  than $q_0$, and the  target halo
can be the satellite ($q>1$) or the main halo ($q<1$).  
Our  interaction  rate  corresponds   to  interaction  rate  \emph{per
  descendant} only in the case $q>1$.

\begin{figure*}
  \begin{center}
    \includegraphics[width=.9\textwidth]{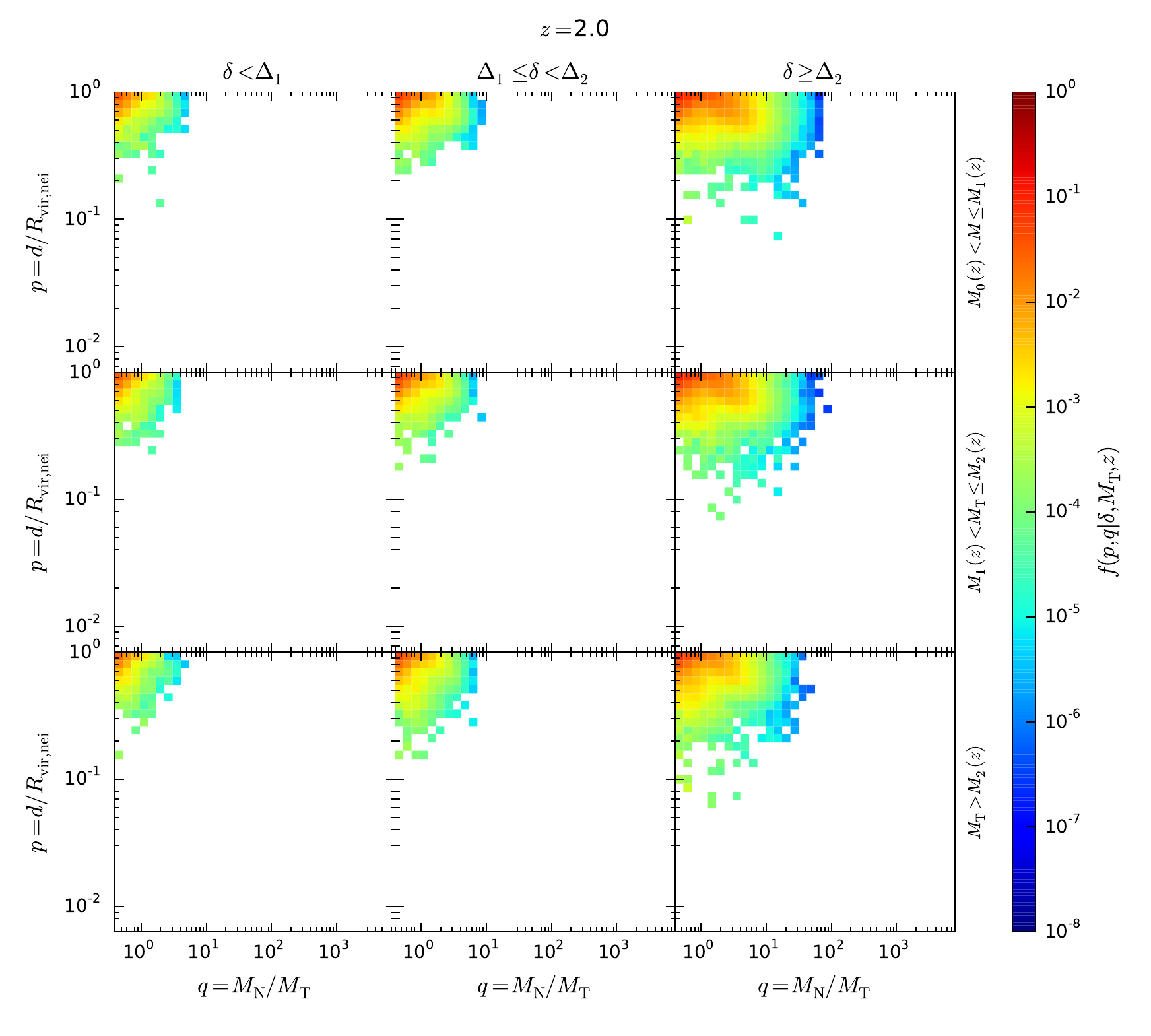}
    \caption{\label{fig:pqz2}%
      Probability  distribution  function $f$  as  a  function of  the
      reduced  distance  $p=d/R_\mathrm{vir,N}$  and  mass  ratio  $q=
      \Mn/\Mt$ at $z=2$.  
      The  three  columns  are low-density  ($\delta  <\Delta_{i,1}$),
      medium-density  ($\Delta_{i,1} \le  \delta <\Delta_{i,2}$),  and
      high-density ($\delta\ge \Delta_{i,2}$) environments.  
      The  three rows  are  low-mass ($M_0<M<M_1$),  intermediate-mass
      ($M_1<M<M_2$), and massive ($\Mt\ge M_2$) target haloes. 
   }
  \end{center}
\end{figure*}

Since our parameter space has 5 dimensions, we divide the whole target
catalogue into subsamples  according to the density,  target mass, and
redshift. We the study the two-dimensional $(p,q)$ distribution 
\begin{equation} 
  \label{eq:f}
  \diff^2 N  = f(p,q|\delta,\Mt,z)\, \diff p\,  \diff q,
\end{equation}
where  $\diff^2 N$  is  the  number of  interactions  in the  interval
$(p,p+\diff  p)$, $(q,q+\diff  q)$ at  fixed redshift  $z$, background
density $\delta$, and halo mass \Mt.

\section{Results} 
\label{sec:res}

\begin{figure*}%[t!]
  \begin{center}
    \includegraphics[width=.9\textwidth]{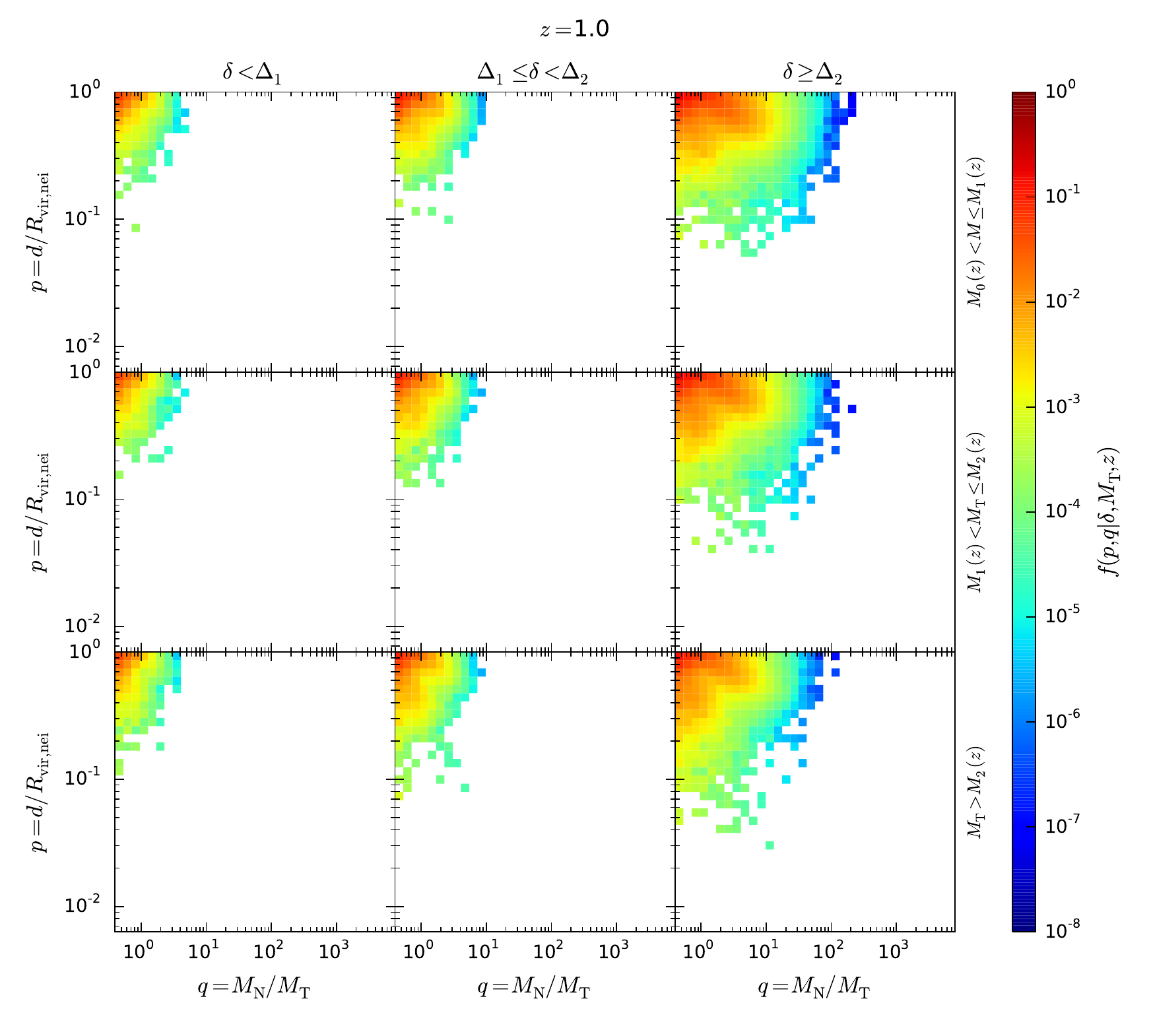}
    \caption{\label{fig:pqz1}%
      Same as Fig.~\ref{fig:pqz2}, at $z=1$.
    }
  \end{center} 
\end{figure*}

\subsection{Mass and density dependence of the interaction rate}
\label{sec:rate}
In this subsection, we are interested in the frequency of interactions
that  fulfil our  definition that  a  target halo  should be  located
within the virial radius of its neighbour.

The  upper  panel  of   Figure~\ref{fig:frho}  shows  the  probability
distribution  of  the  large-scale  density $\delta$  for  all  target
haloes at different redshifts.  
The  distribution  range of  $\delta$  becomes  wider with  decreasing
redshift,  reflecting   the  cosmic   evolution  of   the  large-scale
structures.  
The distribution peaks at $\delta$ of a few units, and the peak height
slightly diminishes with redshift.

The middle panel shows the  fraction $\Gamma(\delta)$ of target haloes
undergoing  interactions  as a  function  of  the large-scale  density
$\delta$.  
In low-density  regions, and  at all redshift,  this fraction  is very
low,  while in  high density  regions ($\delta  > 100$),  the fraction
monotonously rises  to reach a  maximum at $1+\delta \simeq  500$, and
then drops again.  
However, at these  very-high densities, the number of  targets is much
lower, thus the statistical significance of the drop decreases.  
The interaction  rate at  a fixed  density dramatically  increase from
$z=5$ to 2, and then shows little time evolution after this period. 

Finally, the  bottom panel of  the figure  shows the actual  number of
interactions.  
For all $\delta$, the number of interactions increases with decreasing
redshift. 
We  find that  halo-halo  interactions occur  most  frequently in  the
moderately high-density  regions with a  very narrow density  range of
$10<\delta<30$ at all studied redshifts.
Therefore,  even though  halo interactions  occur in  a wide  range of
density environments,  the majority  of interacting haloes  are always
located in the regions with $\delta \simeq 20$.  
This     corresponds      to     haloes     in      filaments     (see
Fig.~\ref{fig:halos_lss}).

\begin{figure*}%[t!]
  \begin{center}
    \includegraphics[width=.9\textwidth]{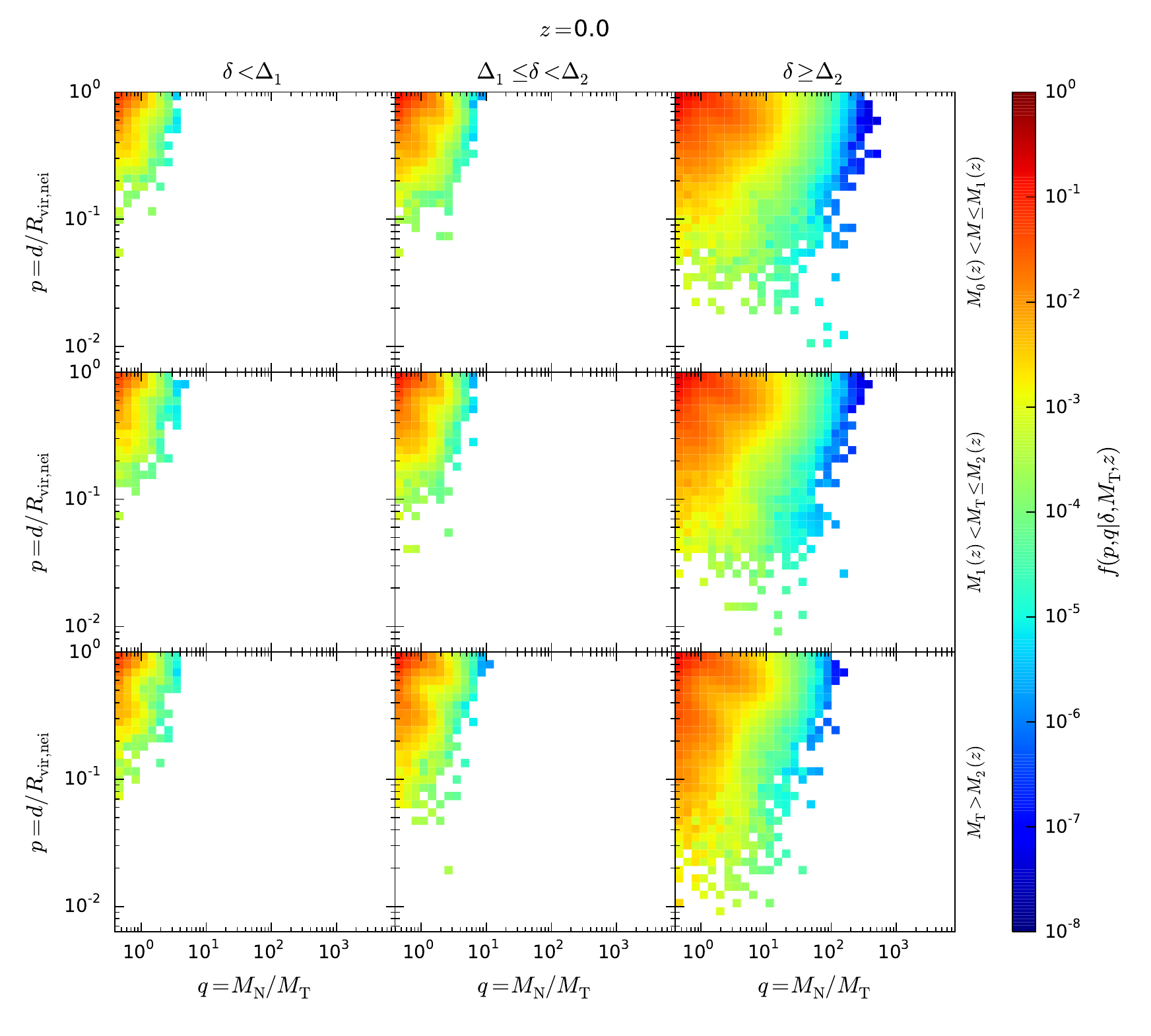}
    \caption{\label{fig:pqz0}%
      Same as Fig.~\ref{fig:pqz2}, $z=0$.
    }
  \end{center} 
\end{figure*}

\bigskip

Taking  advantage of  our unprecedented  catalogue, we  can study  the
interaction rate  as a function  of the mass $M$,  background density,
and redshift separately.  

The  upper-left  panel of  Fig.~\ref{fig:mrho}\subref{fig:mrhoz0}  and
\subref{fig:mrhoz1} show the distribution function of target haloes at
$z=0$ and 1 as a function of the mass and large-scale density.  
This  can  be  seen  as  an  extension  of  the  mass  function  on  a
two-dimensional plane.  
Most targets are located at intermediate densities and low masses. 
At  low densities,  only low-mass  haloes are  found, while  at higher
densities, the range of masses increases. 
It should be noted  that the mass function has a  double peak at fixed
density. 
At very high $\delta$, however, the mass function is almost flat. 
The trend of increasing target mass  with increasing density is due to
host  haloes,  that  follow  the large-scale  environment,  while  the
wide  range of  masses at  a  given density  is due  to the  subhaloes.
Even though  the definitions of  environment are different, we  find a
qualitatively very similar distribution to \citet{2015arXiv150105520Z}
and \citet{2009MNRAS.394.1825F}.  
The former studied  the two-dimensional distribution of  haloes in the
density--$V_\text{max}$    plane,   where    the   maximum    velocity
$V_\text{max}$ is a proxy for the halo mass. 
Our correlation between mass and density agree with their results.
In  the  latter  case,  our  results are  closer  to  the  left  panel
($\delta_7$), that also includes the mass of the target halo.  

The  upper-right panel  in each  subfigure shows  the distribution  of
interacting targets.  
Its  shape  follows  closely  that  of the  targets,  except  for  the
high-mass end at each density.  
This is because of our definition of interactions: 
the  most  massive  haloes  without a  comparable  neighbour  are  not
considered to be interacting.  
Therefore, the double peaks in the mass function disappears.  

The lower-left  panels of  Fig.~\ref{fig:mrho}\subref{fig:mrhoz0} and
\subref{fig:mrhoz1} 
show the interaction rate, defined as  the fraction of target that are
interacting for a given mass and large-scale density.  
It is  by definition proportional to  the ratio of the  upper-right to
the upper-left panels.  
The  rate at  a given  mass  increases with  increasing $\delta$,  and
reaches a plateau at high densities.  
At fixed density, the interaction  rate is constant then drops towards
large masses.  

At higher redshifts,  the distributions look very  similar, but extend
to lower masses and densities.

We  proceeded  in  the  following  way to  fit  the  interaction  rate
$\Gamma(M|\delta,z)$.  
At fixed $(\delta,z)$, we used an empirical function defined by
\begin{equation}
  \label{eq:gamma_m}
  \Gamma(M|\delta,z) = A_0\,\mathrm{erfc}\left(b\log_{10}\left(\frac M
      {M_*}\right)\right), 
\end{equation}
where  $A_0,  M_*$, and  $b$  are  free  parameters depending  on  the
large-scale density $\delta$ and the redshift.
These parameters respectively describe the total interaction fraction,
the characteristic mass at which the  fraction drops, and the speed of
the drop.  
This  function was  selected  to reflect  the  plateau--drop shape  of
$\Gamma(M|\delta)$. 
We remove  the lower and upper  density bins that are  almost empty of
interactions, and for which the statistics are very poor.

We then show the resulting two-dimensional fit of $\Gamma(M,\delta)$ in
the  lower-right panel  of Fig.~\ref{fig:mrho}\subref{fig:mrhoz0}  and
\subref{fig:mrhoz1}.  
The fit is  overall satisfactory, with an underestimation  of the rate
in the high-$\delta$ end of the range.

Figure \ref{fig:params}  show the  dependence of the  parameters $M_*,
A_0$, and $b$ on the large-scale density at each redshift. 
We fit  them with a second-order  polynomial in $\log_{10}(1+\delta)$,
and show it as a dotted line.

\subsection{Time evolution of the interaction rate}
\label{sec:time}

In this  section, we focus  on the  time evolution of  the interaction
rate. 
In order to follow the same  biased objects in a statistical sense, we
proceeded as follows.
At  a  fixed redshift,  we  divided  our  sample  of target  halo  into
subsamples  of equal  number of  targets, constant  at all  redshifts,
according to their mass. 
Then, we divided each mass-subsample into three subsamples according to
the large-scale density of the targets.
In practice, at $z=4$, we selected all target haloes more massive than
$M_2(z=4)  \equiv  \hMsun{8.67e11}$,   corresponding  to  $N_0  \simeq
\num{4.9e6}$.  
We then divided this sample into  three subsamples of the same size in
terms   of   the  number   of   target   halos,  namely,   $\delta   <
\Delta_{2,1}(z=4)$, $\Delta_{2,1}(z=4) <  \delta < \Delta_{2,2}(z=4)$,
and $\delta > \Delta_{2,2}(z=4)$.  
At $z=3.1$, we  found $M_2(z=3.1)$ and $M_1(z=3.1)$ such  that we have
two subsamples according  to the target mass, with the  same number of
targets as at $z=4$.
We then  divided them  into three subsamples  according to  the target
large-scale density.
At lower  redshift, we  introduced another  subsamples at  low masses,
$M_0(z) < M < M_1(z)$.
This  ensures that  the number  of  targets in  each bin  of mass  and
density  is  constant,  about  1.6  millions, and  thus  that  we  are
following statistically similar haloes across different redshifts. 
The redshift-evolution  of $M_i$  and $\Delta_{j,i},  (i\in \{0,1,2\},
j\in\{1,2\})$ is shown in Fig.~\ref{fig:m0m1m2}.

Figure~\ref{fig:Gamma_z}   shows   the   redshift-evolution   of   the
interaction  rate  for  low-  (top),  intermediate-  (middle),  and
high-masses (bottom panel). 
The low-,  intermediate-, and  high-density bins  are shown  as stars,
diamonds, and squares.  
For visualisation sake, we normalised the rates to be 1 at $a=1$.  
We fit the interaction rate as follow:
\begin{equation} 
  \label{eq:gamma_a}
  \Gamma(a) = B\,\text{exp}\left(\left(-\frac{1-a}A\right)^\gamma\right),
\end{equation}
and the values  of $A, B$, and  $\gamma$ in each bin are  shown in the
figure. 
The resulting fit is shown as a dotted line for each subsample.
$\Gamma$        has        an         inflexion        point        at
$ a = 1 - A ( (\gamma-1) / \gamma ) ^ {1 / \gamma}$, which may be seen
at $ a \simeq 0.3$ -- 0.4.  
It is worth noting that this fitting function is able to reproduce the
behaviour of the interaction rate from  $a=0.2$ to 1 and for all range
of mass and density considered.
Note that  the fitting  function is  still positive  at $a=0$,  so the
description is not perfect at very low $a$, but is good enough for the
range of redshifts of interest. 
The actual  value of the interaction  rates can be directly  read from
$B$.  
The shapes are very similar,  rising then reaching saturation, but the
mass and density dependencies can be seen.  
At fixed mass,  the slope is steeper in the  high-density bin, and the
rate in the low-density bin is flatter, showing an earlier saturation.  
The final interaction rate significantly ($B$) grows with the density,
while $A$ slowly decreases with increasing density. 
At fixed density bin, the  slope becomes shallower and the interaction
rate drops toward larger masses.  
This counter-intuitive point  was already seen in the  right panels of
Fig.~\ref{fig:mrho}, where the interaction rate drops at large masses,
and is due to our definition of interactions.  
Also, the raise is somewhat delayed in the largest mass bin. 
$a_{1/2}$,  the time  when the  interaction rate  reaches half  of its
final value, is larger in this bin.

\section{Type of interactions}
\label{sec:type}

After  studying  the effects  of  the  mass  and environments  on  the
interaction  rate, we  study  further to  investigate the  
distributions of $p$ and $q$.

\subsection{Distribution of the distance and mass ratio of interactions} 

Figure  \ref{fig:pqz2}, \ref{fig:pqz1},  and  \ref{fig:pqz0} show  the
distribution $f(p,q|\delta,\Mt,z)$  at $z=2$,  1, and  0, in  the same
bins of density and target halo mass as in \S~\ref{sec:time}.  
The function is normalised in each bin such that 
\begin{equation} 
  \int\!  f(p,q) \, \diff p \, \diff q = \frac { N_\text {inter, bin} }
  { N_\text {target, bin}} 
\end{equation}
over the integration domain, where $N_\text{inter, bin}$ and
$N_\text{target, bin}$ are  the number of interactions  and targets in
each bin.  
Note that by construction, $N_\text{target,  bin}$ is constant in each
panel.
The distribution function  is thus normalised to  the interaction rate
in the bin, as defined in \S~\ref{sec:res} and Fig.~\ref{fig:mrho}.
There is  a clear dependence  on the halo  mass and on  the background
density.

One can also  see in the lower  right panel the existence  of two main
branches: a vertical branch at $q\lesssim 1$ and an oblique branch. 
They correspond to different stages of interactions:
\begin{enumerate}[(i)]
\item The vertical branch, with a large range of $p$ for $q\simeq 1$,  
\item  The oblique  branch,  where  the mass  ratio  increases as  the
  separation becomes smaller.  
  This branch can  be understood as tidal stripping  as the satellites
  orbits  closer   and  closer  to   the  centre  of  the   main  halo
  \citep[e.g.,][]{2007MNRAS.379.1464S}. 
\end{enumerate}
Interestingly, the  target mass and  background density have  a strong
influence on the relative importance of both branches.

At fixed  redshift and mass bin,  increasing the density has  two main
effects.  
First, the range of $q$ becomes larger at higher densities. 
This reflects the fact that massive haloes are located in high-density
regions.
The second effect,  that directly follows from the first  one, is that
branch (ii) is shifted towards larger values of $q$.

When the redshift and the density bins are fixed, going towards higher
masses increases the importance of branch (i).
Moreover, since  the target  becomes more  massive, its  virial radius
increases, and the minimal value of $p$ decreases.

\begin{figure}
  \centering
  \includegraphics[width=\columnwidth]{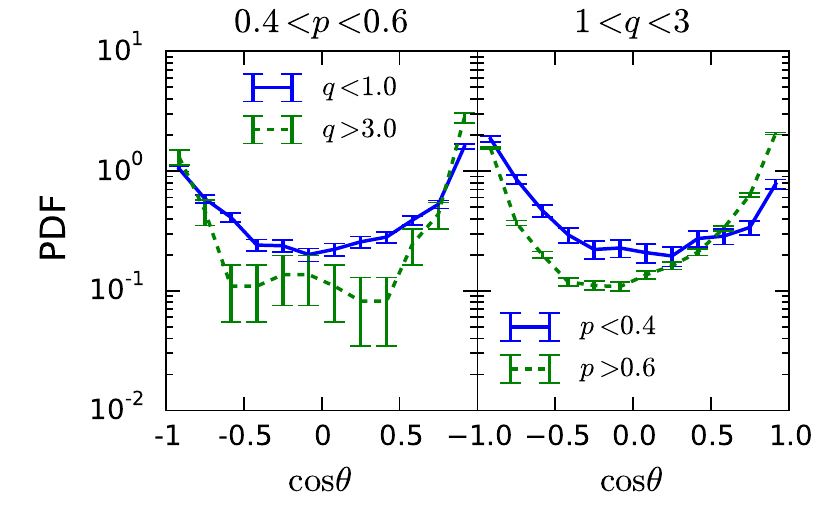}
  \caption{\label{fig:test_dyn_costh}%
    Distribution of $\cos\theta$ for the different subsamples.
    $\theta$ is the angle of the neighbour's velocity vector from the 
    position vector from the target. 
  }
\end{figure}

Finally,  at fixed  mass and  density bin,  there is  also a  redshift
dependence. 
Since  haloes become  more massive,  the  total range  of $q$  becomes
larger.  
The importance  of branch  (ii) fades  away with  decreasing redshift,
while branch (i) becomes more and more important.  

\begin{figure*}%[t!]
  \begin{center}
    \includegraphics[width=\textwidth]{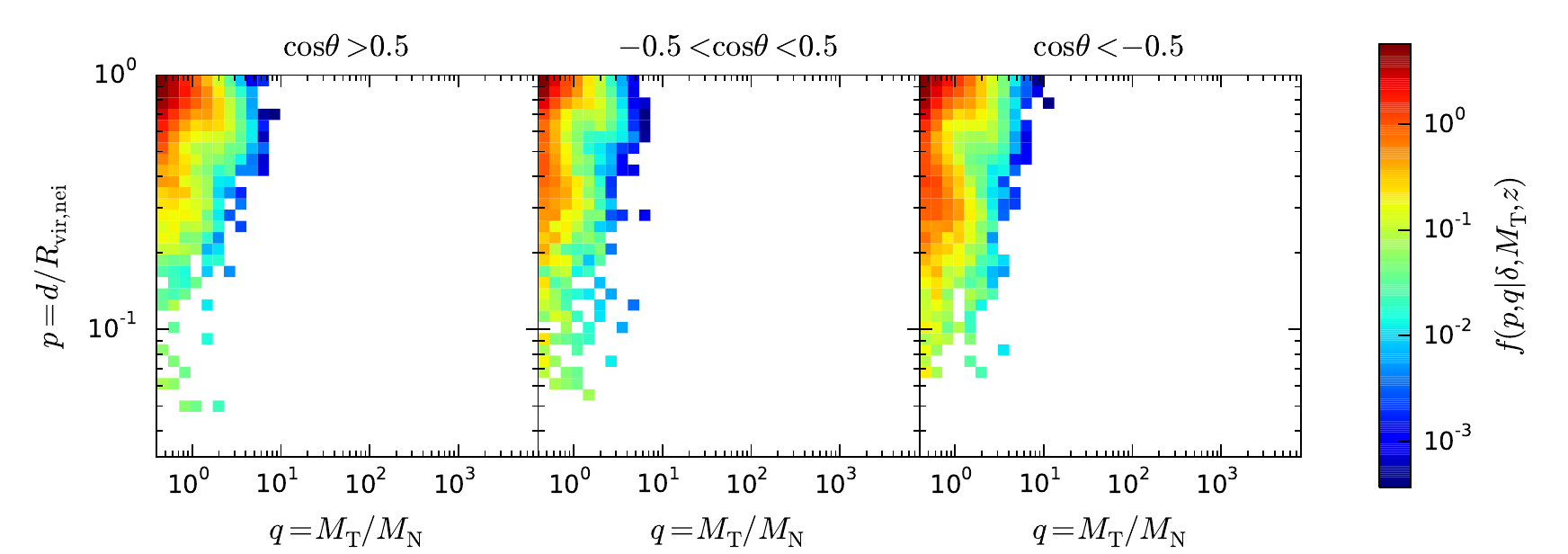}
    \caption{\label{fig:test_dyn_pq}%
      Interaction rate as a function  of $(p,q)$ for interactions with
      $\cos\theta>0.5$  (left),  $-0.5<\cos\theta<0.5$  (middle),  and
      $\cos\theta < -0.5$ (right).  
    }
  \end{center} 
\end{figure*}

We interpret these two branches as follows:
As the interaction starts, the satellite halo (less massive one) enter
within the host, and as it proceeds towards the centre, loses its mass
through tidal stripping (branch (ii)). 
After the closest  approach, the satellite can be  anywhere within the
neighbour, corresponding to branch (i).
In the  next section, we focus  on understanding the physics  of these
two branches.

\subsection{A closer look at the branches}

In  order to  better  understand  the physical  meaning  of these  two
branches, we now focus on the intermediate-density, high-mass panel of
Fig~\ref{fig:pqz0}.  
We  consider  all  interactions  with $0.4<p<0.6$  and  separate  then
between low-  ($q<1$, the  target is  the host  of its  neighbour) and
high-$q$ values ($q>3$), and now  focus on $\theta$, the angle between
the position  and the velocity  of the  neighbour with respect  to the
target, for the two subsamples.  
Tangential orbits have $\cos\theta  = 0$, while receding (approaching)
radial orbits have $\cos\theta = (-)1$. 

The left panel of Fig.~\ref{fig:test_dyn_costh} shows the distribution
of $\cos\theta$.
For a  pure random  case, the distribution  of $\cos\theta$  should be
flat. 
Interactions with  $q<1$ have a flatter  distribution of $\cos\theta$,
reflecting  a  larger amount  of  tangential  orbits than  those  with
$q>3$. 
It can be seen that  the distribution of $\cos\theta$ for interactions
with $q<1$ is flatter, hence more random.
Therefore, the branch at the left of the $p-q$ plane corresponds to the
haloes whose orbits are randomised after they envounter and orbit around
their neighbour.
On the other hand, the interactions with $q>3$ have significantly fewer
tangential orbits compared to the random case.
This means that the upper right branch in the $p-q$ plane is composed of 
haloes in the course of the first encounter in radial orbits.

The right  panel of  Fig.~\ref{fig:test_dyn_costh} shows, for  a fixed
range of neighbour-to-target mass  ratio of $1<q<3$, the distributions
for $p<0.4$ ({solid line}) and $p> 0.6$ ({dotted line}). 
Interaction with $p<0.4$, which belong to the left branch in the $p-q$
plane, show more consistency with random orbits.  
The interactions  with $p> 0.6$  show relative scarcity  of tangential
orbits with  some excess  of receding  radial orbits  over approaching
orbits. 
These observations reinforces  the conclusion that the  left branch is
formed by relaxed haloes.

Figure~\ref{fig:test_dyn_pq}  shows $f(p,q|z=0,  \Delta_{1,2}<\delta <
\Delta_{2,2},  M>  M_2)$,  the  middle  panel  of  the  third  row  of
Fig.~\ref{fig:pqz0},        for        $\cos\theta>0.5$        (left),
$-0.5<\cos\theta<0.5$ (middle), and $\cos\theta < -0.5$ (right).  
One can see that the left  and right panels, corresponding to receding
and  approaching radial  orbits respectively,  have a  prominent upper
right  branch,  while  the  middle row,  corresponding  to  tangential
orbits, shows fewer interactions in this locus. 
This  again reflects  the fact  that  branch (ii)  consists of  haloes
undergoing their  first encounter, with  a rather radial  orbit, while
branch (i) consists of haloes that have already experienced many close
encounters.  

\section{Discussion}
\label{sec:disc}

\subsection{Choice of the neighbour}
\label{sec:neighbour}
In  this section,  we  study  the influence  of  the  choice of  the
neighbour to define interactions.
\citet{2008ApJ...674..784P}  tested the  neighbour with  the largest
tidal energy deposit  (LTED) as the most  influential neighbour, and
found essentially no difference.  
For galaxies in  the field, this makes very  little difference since
the closest  neighbour is more likely  to be the only  one affecting
the target. 
The situation  may be different  in high-density regions,  where two
satellite may be interacting while the main halo, even more distant,
may have a stronger influence \citep{2013MNRAS.436.1765M}. 
\citet{2009ApJ...699.1595P}   showed  that   while  the   structural
parameters  (e.g.,  concentration)  depends  on  the  clustercentric
distance,  the star  formation history  is still  a function  of the
closest neighbour. 

Following  \citet{2008ApJ...674..784P},   we  computed   the  energy
deposit by each of the 20 neighbours used to compute the large-scale
density, provided that it fulfils the conditions ($\Mn > q_0\Mt$ and
$r < R_\text{vir,N}$) using \citet{2008gady.book.....B}: 
\begin{equation}
  \Delta   E   \propto   \frac{\Mn^2    a^2}{\Delta   v^2   (a^2   +
    R_\text{vir,N}^2)^2},
\end{equation}
where $a$ is the rms radius of the target halo, and $\Delta v$ the 
relative velocity.
Note  that we  used the  virial  radius of  the target  $R_\text{T}$
rather  than  the rms  radius,  but  we  expect  the results  to  be
unchanged.  
We then  defined the neighbour  with the  largest $\Delta E$  as the
most influential neighbour.

\begin{figure}
  \includegraphics[width=\columnwidth]{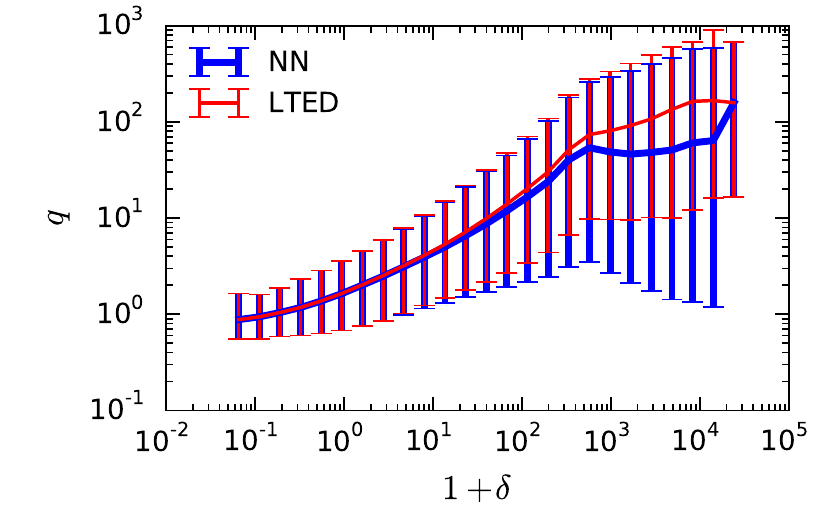}
  \caption{%
    \label{fig:nei}%
    Effect of the definition of the neighbour.
    Shown are the  median value and the 15th and  85th percentiles of
    the mass ratio $q$ for each  density bin for the default nearest
    neighbour case (thick blue lines, NN), and for the maximal energy
    deposit (red line, LTED).
  }
\end{figure} 

Fig.~\ref{fig:nei} shows  the median of  the mass ratio $q$  in each
bin of density.
Interactions  using the  nearest neighbour  (NN) are  shown in  red,
thick lines, and the LTED case is shown in blue line.  
The error bars show the 15th and 85th percentiles.
As  expected, the  median  of  the mass  ratio  increases at  larger
densities since it is more likely to find a more massive neighbour.
Up  to $\delta  \simeq 500$,  the choice  of the  neighbour has  no
effect.
At higher densities, as expected,  the neighbour with the LTED
is not  necessarily the  NN, but  more likely to  be a  more massive
halo, which leads to larger mass ratios.
Since most interactions occur at rather low densities, $\delta$ of a
few tens, the choice of the neigbhbour has no effect.
However, if one  is interested in interactions  within clusters, one
should define the neigbhour carefully.

\subsection{Comparison with previous work}

Previous        theoretical        work        from        simulations
\citep[e.g.,][]{2008MNRAS.386..577F,       2009MNRAS.394.1825F,
  2010ApJ...719..229G}  studied the  merger rate  per time  unit using
merger  trees,   and  typically   found  a   merger  rate   of  $\diff
{N_\text{merg}/\diff      t}     \propto      (1+z)^\alpha$,     while
\citet{2009MNRAS.394.1825F}    found     an    environmental    factor
$(1+\delta_7)^\beta \left(M /{M_{12}}\right)^\gamma$, where $\delta_7$
is the local density computed over a sphere of radius \hMpc{7}.  
However, our  approach differs from previous  theoretical studies in
that we do  not use merger trees to define  interactions, but rather
look at the distance to the closest neighbour at a given time.  
This makes direct comparison with previous work difficult.  
Indeed,  in our  approach,  the  same interaction  may  be counted  at
different redshifts,  while in other  studies it will only  be counted
when the merger is complete. 

Our approach is by construction closer to observations in that we look
at  the instantaneous  interaction fraction,  namely, the  fraction of
target  that  are  undergoing  an interaction  at  a  given  redshift,
corresponding to $f_\text{m}$ in \citep{2008ApJ...672..177L}.  
However, our  condition that the  target should lie within  the virial
radius of its neighbour,  increases our interaction numbers especially
at low-redshifts where the physical  virial radius of the studied mass
range is typically 0.5 to \hMpc{1}, compared to close-pair counts that
limit the separations to a few \hkpc{10}.  

\smallskip

The main advantage of our approach, besides its simplicity, is that it
does not require observations to use a merger timescale, that can be a
source of uncertainty.
Instead, the observed merger fraction  can be directly compared to our
fitting formulae.
However,  one should  keep in  mind  that, since  we consider  distant
interactions,  our pair  separation  can be  larger  than the  typical
\hkpc{30} separation used to calculate merger rates.

\smallskip

Observationally, most  studies define the interaction  radius, i.e.,
the  largest radius  to  define  an interacting  pair,  as fixed  in
physical scales. 
\citet{2015A&A...576A..53L}  compiled several  study of  close pairs
for  $0<z<1.2$, with  a minimal  projected separation  of \hkpc{100}
(physical), and fitted an interaction fraction of $f = f_0 (1+z)^m$. 
They found  that $f_0$ decreases  and $m$ increases  with decreasing
magnitude cut.  
While  our fitting  formula  are  different, $f_0$  and  $m$ can  be
related to our $B$ and $\gamma$ parameters.
However, their behaviours are quite different, since $B$ is constant
with mass, except in the largest density bin.
The  main  difference with  our  approach  is  the  use of  a  fixed
projected radius for the definition  of interacting pairs, while our
interaction distance depends on the neighbour.
At fixed mass, and in proper units, the virial radius increases with
decreasing  redshift  (since  $R_\text{vir}$  and  $\Delta_\text{c}$
decrease, c.f.  equations~\ref{eq:rvir} and \ref{eq:delta_c}).
Moreover,  haloes also  accrete mass,  which also  makes the  virial
radius grow.
These  considerations results  in a  larger interaction  fraction at
lower redshift in our definition.

\smallskip

\section{Summary} 
\label{sec:ccl}

In this study, we used the massive Horizon run 4 cosmological $N$-body
simulation  to investigate  halo  interactions,  by considering  three
aspects  of the  environment,  namely, internal  (target mass),  local
(distance  to  the  nearest neighbour),  and  large-scale  (background
density).  

This simulation has  the largest size among all  simulations that have
been used  to study  galaxy interactions, which  enables to  study the
environment dependence and very  good statistics, while its resolution
is   high   enough    to   study   haloes   with    masses   down   to
$\hMsun{5e11}$.  
We defined  target and  neighbour haloes as  haloes more  massive than
$\num{5e11}$ and  \hMsun{2e11}, and we  define a target  of mass
$\Mt$  to be  \emph{interacting} if  it is  located within  the virial
radius of a neighbour more massive than $0.4 \Mt$.  

Our findings are summarised as follow:
\begin{itemize}
\item 
  Interactions  preferentially   occur  at   intermediate  large-scale
  background  densities,   $\delta  \simeq  20$,  regardless   of  the
  redshift, and for targets more massive than \hMsun{5e11}.
  The interaction  rate, \ie\  the fraction  of targets  undergoing an
  interaction,  reaches its  maximum at  larger values:  $\delta\simeq
  1000$.  
\item 
  We  propose  a new  functional  fit  (eq.~\ref{eq:gamma_m}) for  the
  interaction rate  as a function  of the target mass  and large-scale
  background   mass  density   at   fixed  redshift,   and  give   the
  redshift-evolution of the fit parameters.  
\item  The interaction  rate as  a function  of time  has a  universal
  shape, rising and saturating. 
  We  propose  a  fitting  formula  for  the  time  evolution  of  the
  interaction rate (eq.~\ref{eq:gamma_a}).
  This function strongly depends on the large-scale density and target
  mass.
  Larger density  yield a steeper  slope and larger  final interaction
  rate, and massive haloes saturate later.  
\item We reported the existence of two branches in the two-dimensional
  distribution $f(p,q|\delta,\Mt,z)$,
  namely:  an  oblique branch  where  the  distance to  the  neighbour
  decreases as  the mass ratio  increases, reflecting the loss  of the
  satellite mass through tidal stripping  as it orbits within the main
  halo. 
  A  second branch  at  lower  $p$ for  the  same  $q$ corresponds  to
  interactions  that already  passed  the closest  approach and  whose
  orbits became randomised.
\item The relative importance of the two trends strongly depend on the
  large-scale density, target mass, and redshift.  
\end{itemize}

In a paper  in preparation, we will study the  effects of interactions
on parameters such as the alignment of  the spin and major axes of the
interacting haloes, and their orbits.
In a next step, we will add hydrodynamics to study the galaxy (rather
than the halo) interaction rate,  which can be significantly different
\citep{2010ApJ...724..915H}, and  to probe the effects  of gas physics
on the morphological transformation of galaxies.  

We  argue  that observers  may  use  our  definitions to  measure  the
interaction fraction. 
In a  next paper, we  plan to apply  our method to  publicly available
observational data  in order to  compare the evolution  of interaction
fraction between observations and our theoretical results.  

\subsection*{Acknowledgements}
We thank KIAS Center  for Advanced Computation for providing computing
resources.
We  thank the  anonymous referee  for her/his  precious comments  that
helped improving this manuscript.
\bibliographystyle{mn2e} 
\bibliography{biblio}

% \clearpage

\appendix

\section{$\delta$ and the cosmic web}
\label{sec:delta}

\begin{figure} 
  \centering
  \includegraphics[width=.8\columnwidth]{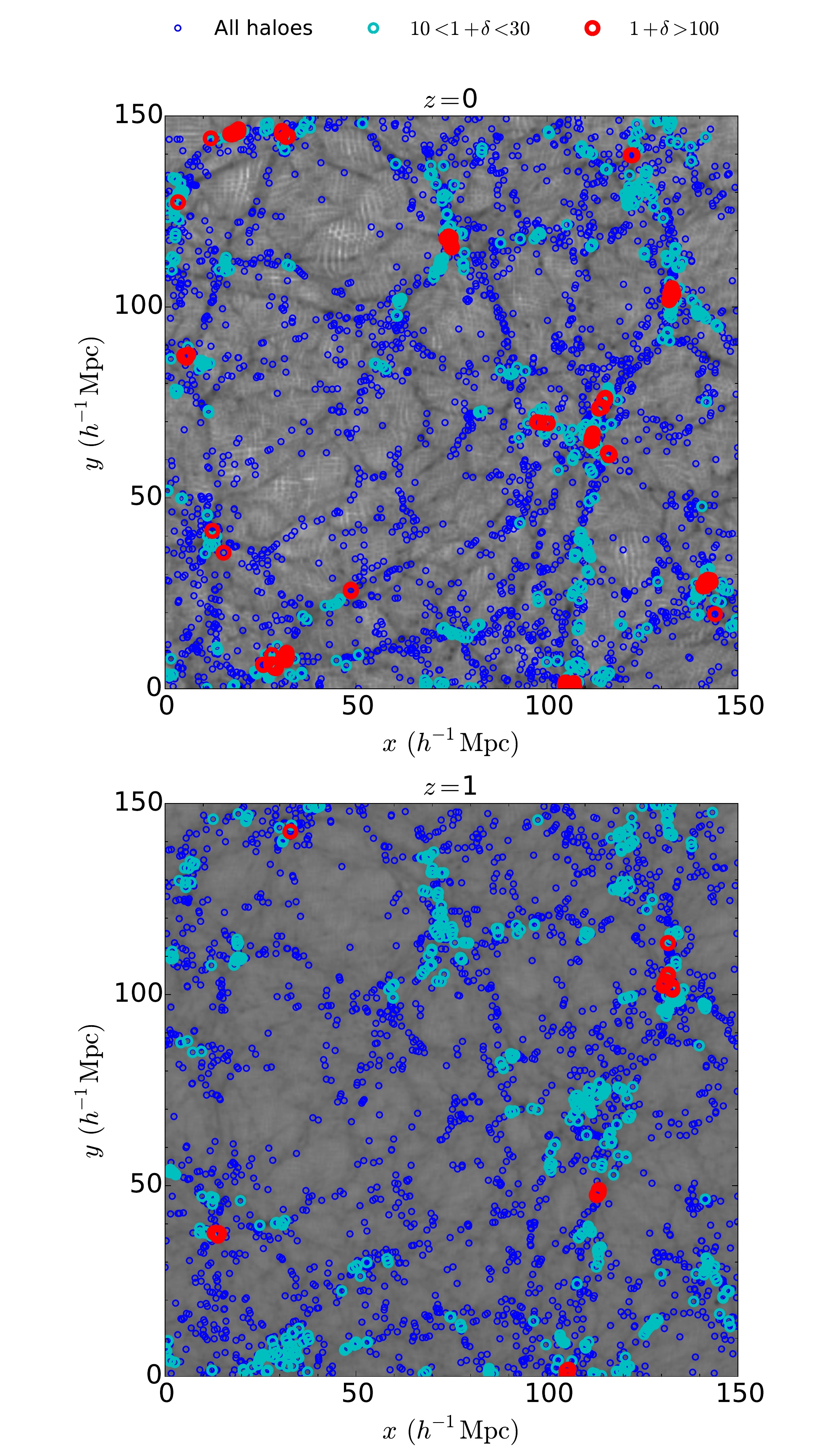}
  \caption{%
    \label{fig:halos_lss}%
    Distribution of haloes with  different overdensities in the cosmic
    web at $z=0$ (top) and 1 (bottom).  
    The grey background is the  density field computed over \hMpc{0.5}
    pixels with a TSC method.  
    The larger values (large  red circles, $1+\delta>100$) are located
    in  the nodes,  while intermediate  valuers (medium,  cyan circles
    $10<1+\delta<30$) are located in the filaments.  
  }
\end{figure}

It is interesting  to understand the relation  between the overdensity
parameter $\delta$  and the  cosmic web, namely,  where haloes  with a
given $\delta$ are located within the cosmic web.
Figure~\ref{fig:halos_lss} shows  a $150 \times 150  \times 20$\hMpc{}
slice at $z=0$ (top) and 1 (bottom).  
High-$\delta$ haloes are shown in large, red circles,
intermediate-$\delta$  ($10<1+\delta<30$)  in  medium,  cyan  circles,
while other haloes are shown in small, blue circles.
Haloes are  overplotted on top  of the  density map computed  with the
Triangular-Shape Clouds, with a pixel  size of \hMpc{0.5}, the same as
used for the PM calculation.
At $z=0$, haloes  with $\delta>100$(red) can be found  in nodes, while
intermediate ($10<\delta<30$) are found in filaments. 
On the other hand, at  $z=1$ the overall density fluctuation amplitude
is lower,  and the haloes  with $\delta>100$  are located only  in the
rarer highest  density nodes. Since  we vary the density  threshold to
cope with the evolution of the  growth factor, the type of environment
corresponding to  each density  bin is supposed  to be  independent of
redshift. 
We conclude  that the overdensity $1+\delta$  is a good proxy  for the
environment.

\section{Effects of $M_0, M_1$, and $M_2$}

\begin{figure*}
  \includegraphics[width=.9\textwidth]{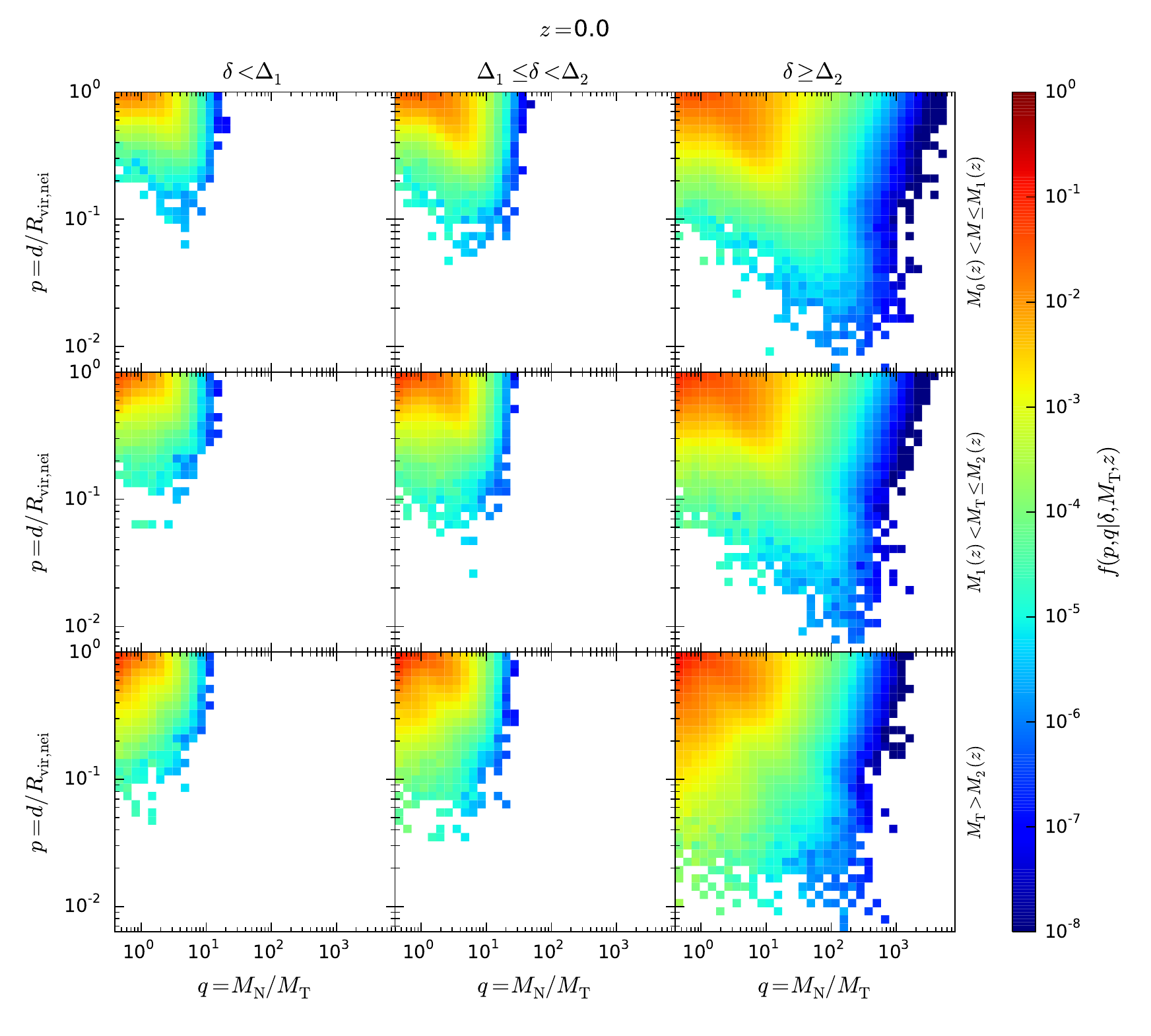}
  \caption{\label{fig:pqz0_zmax2}%
    Same as Fig.~\ref{fig:pqz0}, with $z_\text{max} = 2$. 
  }
\end{figure*}

Fig.~\ref{fig:pqz0_zmax2}  is  the  same as  Fig.~\ref{fig:pqz0},  but
using $z_\text{max}=2$ instead of 4.  
The larger number  of targets available from $z=2$ enables  us to keep
better statistics, but prevents  from doing a redshift-evolution study
as in \S~\ref{sec:time}.
It results in a much higher number of targets in each panel, about 20
millions, from $z=2$ to 0. 

\begin{figure}  
  \includegraphics[width=\columnwidth]{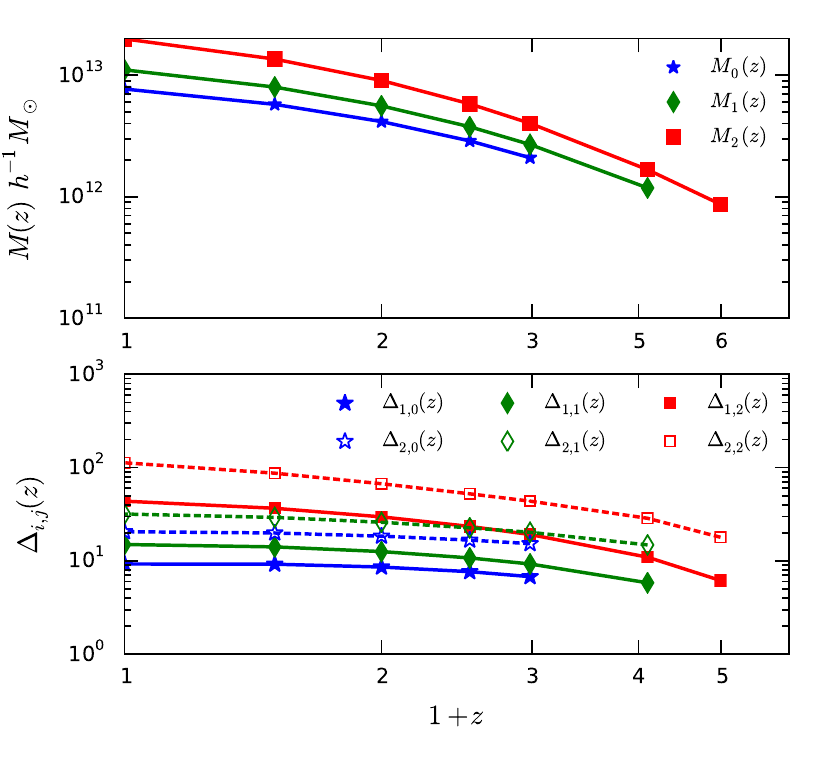}
  \caption{\label{fig:m0m1m2}%
    Redshift evolution  of $M_j$  (top), and  $\Delta_{i,j}$ (bottom),
    where $j$ stands for the mass bin and $i$ for the density bin, for
    the $z_\text{max}=4$ case in \S~\ref{sec:time}.  
  }
\end{figure}

The value  of $M_0,  M_1$, and  $M_2$ are  respectively \num{6.86e11},
\num{1.1e12},   and   \hMsun{2.14e12},  compared   to   \num{7.68e12},
\num{1.10e13}, and \hMsun{1.99e13} for  the $z_\text{max}=4$ (shown in
Fig.~\ref{fig:m0m1m2})  case,  which  implies   that  all  targets  in
Fig.~\ref{fig:pqz0}     are     in      the     lower     panel     of
Fig.~\ref{fig:pqz0_zmax2},  resulting  in  a  more  important  oblique
branch.  

\label{lastpage}

\clearpage

\end{document}